\documentstyle[12pt,epsfig]{article}
\date{\today}

\newcommand{\scs}{\scriptscriptstyle}
\newcommand{\be}{\begin{equation}}
\newcommand{\ee}{\end{equation}}
\newcommand{\bea}{\begin{eqnarray}}
\newcommand{\eea}{\end{eqnarray}}
\newcommand{\f}{\frac}
\def\slash#1{\setbox0=\hbox{$#1$}#1\hskip-\wd0\dimen0=5pt\advance
       \dimen0 by-\ht0\advance\dimen0 by\dp0\lower0.5\dimen0\hbox
         to\wd0{\hss\sl/\/\hss}}
\def\simleq{\stackrel{<}{\scs \sim}}
\def\simgeq{\stackrel{>}{\scs \sim}}

\title{$t\rightarrow cV$ via SUSY FCNC couplings  in the unconstrained MSSM}
\author{Jian Jun Liu$^a$, Chong Sheng Li$^a$\footnote{csli@pku.edu.cn}, Li Lin Yang$^a$ and Li Gang Jin$^b$\\
{\small $^a$ Department of Physics, Peking University, Beijing
100871, China} \\ {\small $^b$ Institute of Theoretical Physics,
Academia Sinica, P. O. Box 2735,
Beijing 100080, China} \\
}
\date{\today}

\begin{document}
\maketitle
\begin{abstract}
We recalculate the branching ratios for $t\rightarrow cV$
($V=g,\gamma,Z$) induced by SUSY FCNC couplings within the general
unconstrained MSSM framework using mass eigenstate approach. Our
results show that the branching ratios for these decays are larger
than ones reported in previous literatures in the MSSM with
R-parity conservation, and they can reach $\sim 10^{-4}$, $\sim
10^{-6}$, and $\sim 10^{-6}$, respectively, for favorable
parameter values allowed by current precise experiments. Thus, the
branching ratios for $t\rightarrow cg$ and $t\rightarrow c\gamma$
may be measurable at the LHC.
\end{abstract}

\vspace{1.5cm} \noindent  Keywords: top quark, MSSM, FCNC


\newcommand{\beq}{\begin{eqnarray}}
\newcommand{\eeq}{\end{eqnarray}}

\def\slash#1{\setbox0=\hbox{$#1$}#1\hskip-\wd0\dimen0=5pt\advance
       \dimen0 by-\ht0\advance\dimen0 by\dp0\lower0.5\dimen0\hbox
         to\wd0{\hss\sl/\/\hss}}
\def\simleq{\stackrel{<}{\scs \sim}}
\def\simgeq{\stackrel{>}{\scs \sim}}

\def\mathrm#1{{\rm #1}}
\def\ra{\rightarrow}
\def\kk{\mbox{$\bar{K}^0K^0$~}}
\def\bb{\mbox{$\bar{B}^0B^0$~}}
\def\dd{\mbox{$\bar{D}^0D^0$~}}
\def\bbd{\mbox{$\bar{B}_d^0B_d^0$~}}
\def\BSG{\mbox{$Br(B\ra X_s\gamma)$~}}
\def\bsg{\mbox{$b\ra s\gamma$~}}
\def\epsk{\mbox{$\epsilon_K$~}}
\def\dmbd{\mbox{$\Delta m_{B_d}$~}}
\def\TeV{\mathrm{TeV}}
\def\GeV{\mathrm{GeV}}

\newpage

\section{Introduction}

The top quark flavor changing neutral current (FCNC) processes
$t\rightarrow cV$ ($V=g,\gamma,Z$) have tiny branching ratios in
the standard model (SM) \cite{eilam}, and are too small to be
measurable in the future colliders, and thus any detected signal
of these rare decay events definitely indicates some new physics
beyond the SM. Actually, $t\rightarrow cV$ ($V=g,\gamma,Z$) have
been studied in various new physics models beyond the SM in
detail, such as the two-Higgs-doublet model (2HDM)
\cite{eilam,goush}, the technicolor model (TC) \cite{yue}, the
top-color-assisted technicolor model (TC2) \cite{lu}, the models
with extra vector-like quark singlets\cite{agu}, the minimal
supersymmetry (SUSY) extension of the SM (MSSM) with R-parity
conservation \cite{csldecay}-\cite{sola} and without R-parity
conservation \cite{yang}. The decay branching ratios for
$t\rightarrow cV$ ($V=g,\gamma,Z$) are enhanced in general several
orders of magnitude in these new physics models. The MSSM, which
is believed as one of the most attractive candidates of new
physics model, has gotten many attentions, and the investigation
of $t\rightarrow cV$ ($V=g,\gamma,Z$) in the MSSM is a long story.
C.S.~Li {\it et al.} studied one-loop SUSY-QCD and SUSY-EW
contributions in Ref.~\cite{csldecay}, subsequently G. Couture
{\it et al.} recalculated and generalized the SUSY-QCD corrections
to include the left-hand (LH) squark mixing in Ref.~\cite{cou1}
and the right-hand (RH) squark mixing in Ref.~\cite{cou2}. All
works above are within the framework of the MSSM with
flavor-universal soft SUSY breaking terms. Later J.L.~Lopez {\it
et al.} further generalized the SUSY-EW corrections to the case of
including neutralino-quark-squark loops in Ref.~\cite{lopez}. and
G.M.~de~Divitiis {\it et al.} reinvestigated $t\rightarrow cV$
($V=g,\gamma,Z$) in the universal case as well as non-universal
case in Ref.~\cite{divit}, and obtained different results from
Ref.~\cite{csldecay,cou1} due to the calculation of the relevant
SUSY mixing angles and diagrams involving a helicity flip in the
gluino line, which was confirmed by J.~Guasch {\it et al.} in a
RG-based framework for $t\rightarrow cg$ decay~\cite{sola}. All
the above results of the MSSM are summarized in Table 1 (for
comparing, we also list the results of the SM and new physics
models mentioned above), one can find that they are all below $5
\times 10^{-5}$, which is the roughly estimated sensitivities for
the measurements of top rare decay at the LHC with $100fb^{-1}$ of
integrated luminosity~\cite{sola}.

\begin{table}[ht]
\begin{center}
\begin{tabular}{cccccccccccccc}
\hline\hline \scriptsize decay & & & & & & & & \multicolumn{6}{c}{MSSM} \\
{\scriptsize  mode} & \raisebox{0.4cm}[0pt]{\scriptsize SM}&
\raisebox{0.4cm}[0pt]{\scriptsize 2HDM}&
\raisebox{0.4cm}[0pt]{\scriptsize TC}&
\raisebox{0.4cm}[0pt]{\scriptsize TC2}&
\raisebox{0.4cm}[0pt]{\scriptsize CKM2}&
\raisebox{0.4cm}[0pt]{\scriptsize CKM1}&
\raisebox{0.4cm}[0pt]{\scriptsize RPV}
 &\scriptsize \cite{csldecay} &\scriptsize \cite{cou1} &\scriptsize \cite{cou2} &\scriptsize ~\cite{lopez} &\scriptsize \cite{divit} &\scriptsize \cite{sola} \\
\hline
\scriptsize$t\rightarrow cg$ &\scriptsize $10^{-13}$ &\scriptsize $10^{-5}$&\scriptsize $10^{-6}$&\scriptsize $10^{-5}$&\scriptsize $10^{-11}$&\scriptsize $10^{-7}$&\scriptsize $10^{-3}$&\scriptsize $10^{-6}$ &\scriptsize $10^{-5}$&\scriptsize$10^{-5}$ &\scriptsize$10^{-5}$&\scriptsize$10^{-5}$&\scriptsize $10^{-6}$\\
\scriptsize$t\rightarrow c\gamma$ &\scriptsize $10^{-14}$ &\scriptsize $10^{-7}$&\scriptsize $10^{-8}$&\scriptsize $10^{-7}$&\scriptsize $10^{-12}$&\scriptsize $10^{-8}$&\scriptsize $10^{-5}$&\scriptsize $10^{-8}$ &\scriptsize $10^{-7}$ &\scriptsize $10^{-7}$&\scriptsize $10^{-7}$  &\scriptsize $10^{-6}$& \\
\scriptsize$t\rightarrow cZ$ &\scriptsize $10^{-15}$ &\scriptsize $10^{-6}$ &\scriptsize $10^{-7}$&\scriptsize $10^{-5}$&\scriptsize - &\scriptsize $10^{-4}$&\scriptsize $10^{-4}$&\scriptsize$10^{-8}$ &\scriptsize $10^{-6}$&\scriptsize $10^{-6}$&\scriptsize $10^{-7}$  &\scriptsize $10^{-6}$ &  \\
\hline\hline
\end{tabular}
\caption{$t\rightarrow cV$ ($V=g,\gamma,Z$) branching ratios of
previous calculations. 'CKM1' and 'CKM2' refer to models with
extra vector-like up-type quark singlets and down-type quark
singlets, respectively \cite{agu}, and 'RPV' refers to SUSY models
allowing R-parity violation.}
\end{center}
\end{table}
However, all the previous works are limited to some constrained
MSSM, in which some strong assumptions or additional parameters
besides ones in the MSSM are introduced to describe the FCNC
couplings, but no any strong theoretical reasons of them have been
found so far. It is necessary to study the FCNC top quark decays
in the unconstrained MSSM \cite{rosiek}, where the assumptions
about the soft SUSY breaking terms are relaxed and new sources of
flavor violation are presented in the mass matrices of sfermions,
and consequently, some large contributions to FCNC processes
induced by SUSY FCNC couplings (neutralino-quark-squark coupling
and gluino-quark-squark coupling) can be obtained. Since the
contributions to the top FCNC decays mediated by the charged
current interactions (from $W^\pm$, $H^\pm$, $G^\pm$ and
$\tilde{\chi}^\pm$) are invisibly small as shown in the previous
works \cite{csldecay}-\cite{sola} and can not be enhanced in this
framework, in this paper we will reinvestigate the $t\rightarrow
cV$ ($V=g,\gamma,Z$) only via SUSY FCNC couplings in the
unconstrained MSSM, and try to show what are the maximal branching
ratios for $t\rightarrow cV$ ($V=g,\gamma,Z$) in the MSSM using
SUSY parameters allowed by current data, and whether they can be
detected at the LHC.

\section{\label{sec:pro}The $t\rightarrow cV \ (V=g,\gamma,Z)$ process induced by
SUSY FCNC} In the super-CKM basis \cite{rosiek}, in which the mass
matrices of the quark fields are diagonal by rotating the
superfields, the up squark mass matrix ${\cal M}_{\tilde{U}}^2$ is
a $6 \times 6$ matrix, which has the form:
\begin{equation}
\left( \begin{array}{cc}
  \left(M^2_{\,{\tilde{U}}}\right)_{LL}+(m^2_{u}\+\cos 2\beta \, m_Z^2
   \left(\frac{1}{2}-\frac{2}{3}\sin^2\theta_W \right))
{{\mathchoice {\rm 1\mskip-4mu l} {\rm 1\mskip-4mu l} {\rm
1\mskip-4.5mu l} {\rm 1\mskip-5mu l}}}_3\           &
  \left(M^2_{\,{\tilde{U}}}\right)_{LR}-\mu(m_{u}\cot\beta)\,{{\mathchoice
{\rm 1\mskip-4mu l} {\rm 1\mskip-4mu l} {\rm 1\mskip-4.5mu l} {\rm
1\mskip-5mu l}}}_3 \
                                                     \\[1.01ex]
 \left(M^2_{\,{\tilde{U}}}\right)_{LR}^{\dagger}-\mu (m_{u}\cot\beta)\,{{\mathchoice
{\rm 1\mskip-4mu l} {\rm 1\mskip-4mu l} {\rm 1\mskip-4.5mu l} {\rm
1\mskip-5mu l}}}_3 \ &
             \left(M^2_{\,{\tilde{U}}}\right)_{RR}+(m^2_{u}\+\cos2\beta\,m_Z^2
   \left(\frac{1}{2}-\frac{2}{3}\sin^2\theta_W \right))
{{\mathchoice {\rm 1\mskip-4mu l} {\rm 1\mskip-4mu l} {\rm
1\mskip-4.5mu l} {\rm 1\mskip-5mu l}}}_3\
 \end{array} \right)  \nonumber \end{equation} where $\theta_W$ is
the Weinberg angle, ${{\mathchoice {\rm 1\mskip-4mu l} {\rm
1\mskip-4mu l} {\rm 1\mskip-4.5mu l} {\rm 1\mskip-5mu l}}}_3$
stands for the $3 \times 3$ unit matrix, the angle $\beta$ is
defined by $\tan \beta \equiv v_2/v_1$, the ratio of vacuum
expectation values of the two Higgs doublets, $\mu$ is the Higgs
mixing parameter in the superpotential, and
$(M^2_{\tilde{U}})_{LL}$, $(M^2_{\tilde{U}})_{RR}$, and
$(M^2_{\tilde{U}})_{LR}$ contain the flavor-changing entries,
which are given by \bea
\begin{array}{ccc}
(M^2_{\tilde{U}})_{LL} = V_L^U M^2_Q V_L^{U\dagger},
\hspace{0.4cm}& (M^2_{\tilde{U}})_{RR} = V_R^U (M^{2}_U)^T
V_R^{U\dagger}, \hspace{0.4cm}& (M^2_{\tilde{U}})_{LR} = -
\mbox{\Large $\f{v \sin \beta}{\sqrt{2}}$} V_L^U A_U^{\ast}
V_R^{U\dagger},
\end{array}
\eea respectively. Here $M^2_Q$, $M^{2}_{U,D}$ and $A_{U,D}$ are
the soft broken $SU(2)$ doublet squark mass squared matrix, the
$SU(2)$ singlet squark mass squared matrix and the trilinear
coupling matrix, respectively. They are directly related to the
mechanism of SUSY breaking, and are in general not diagonal in the
super-CKM basis. Furthermore, $(M^2_{\tilde{U}})_{LR}$, arising
from the trilinear terms in the soft potential, namely $A_{U,ij}
H_U \,{\widetilde{U}}_i {\widetilde{U}}_j^{c}$, is not hermitian.
The matrix ${\cal M}^2_{\tilde{U}}$ can further be diagonalized by
an additional $6\times 6$ unitary matrix $Z_U$ to give the up
squark mass eigenvalues \bea \left({\cal
M}^2_{\tilde{U}}\right)^{diag} = Z_U^{\dagger} {\cal
M}^2_{\tilde{U}} Z_U \label{eq:zudef}. \eea

Thus, we get new sources of flavor-changing neutral current:
neutralino-quark-squark coupling and gluino-quark-squark coupling,
which arise from the off-diagonal elements of
$(M^2_{\tilde{U}})_{LL}$, $(M^2_{\tilde{U}})_{LR}$ and
$(M^2_{\tilde{U}})_{RR}$, and can be written as
($I=1,2,3$,$i=1,...,6$, $j=1,2,3,4$)
\begin{eqnarray} \tilde{g}^a-\tilde{q}_{ir}-q_{Is}: &&\quad
i\sqrt{2}g_sT^a_{rs}
\left[-(Z_{U})_{Ii}P_L + (Z_{U})_{(I+3)i} P_R\right]\nonumber \\
 \tilde{\chi}^0_j-\tilde{q}_{ir}-q_{Is}: &&\quad
i\delta_{rs}\{[\frac{-e}{\sqrt{2}s_Wc_W}(Z_{U})_{Ii}(\frac{1}{3}s_W(Z_N)_{1j}+c_W(Z_N)_{2j})
-Y_u^I(Z_U)_{(I+3)i}(Z_N)_{4j}]P_L\nonumber\\
&&+[\frac{2\sqrt{2}e}{3c_W}(Z_U)_{(I+3)i}(Z_N)_{1j}-Y_u^I(Z_U)_{Ii}(Z_N)_{4j}]P_R\}\nonumber.
\end{eqnarray}
Here $s_W\equiv \sin\theta_W$, $c_W\equiv \cos\theta_W$,
$P_{L,R}\equiv (1\mp \gamma_5)/2$, $T^a_{rs}$ is the $SU(3)$ color
matrix with color index a, r, s, and the unitary transformation
$Z_N$ diagonalizes mass matrix of gauginos and higgsinos to obtain
the neutralinos. Thus the flavor changing effects of soft broken
terms $M^2_{Q}$, $M^2_{U}$ and $A_{U}$ on the observables can be
obtained through the matrix $Z_U$.

For the aim of this paper, the following strategy in the numerical
calculations of the decay branching ratios of $t\rightarrow cV$
will be used: first we deal with the LL, LR, RL and RR blocks of
the matrix ${\cal M}^2_{\tilde{U}}$ separately and in each block
we only consider the effects of individual element on the top
quark rare decays, and then we investigate the interference
effects between some different entries within one block and the
interference effects between different blocks. In order to
simplify the calculation we further assume that all diagonal
entries in $(M^2_{\tilde{U}})_{LL}$, $(M^2_{\tilde{U}})_{LR}$,
$(M^2_{\tilde{U}})_{RL}$ and $(M^2_{\tilde{U}})_{RR}$ are set to
be equal to the common value $M^2_{\rm{SUSY}}$, and then normalize
the off-diagonal elements to $M^2_{\rm{SUSY}}$ \cite{besmer,gam},
\begin{eqnarray}
&& (\delta_{U}^{ij})_{LL} =
\frac{(M^2_{\tilde{U}})_{LL}^{ij}}{M^2_{\rm{SUSY}}}\,,
\hspace{1.0truecm} (\delta_{U}^{ij})_{RR} =
\frac{(M^2_{\tilde{U}})_{RR}^{ij}}{M^2_{\rm{SUSY}}}\,,
\hspace{1.0truecm} \nonumber \\
&& (\delta_{U}^{ij})_{LR} =
\frac{(M^2_{\tilde{U}})_{LR}^{ij}}{M^2_{\rm{SUSY}}}\,,
\hspace{1.0truecm} (\delta_{U}^{ij})_{RL} =
\frac{(M^2_{\tilde{U}})_{RL}^{ij}}{M^2_{\rm{SUSY}}}\,,
\hspace{1.0truecm} (i \ne j,i,j=1,2,3). \label{deltadefb}
\end{eqnarray}
Thus $(M^2_{\tilde{U}})_{LL}$ can be written as follows: \be
(M^2_{\tilde{U}})_{LL} =M^2_{\rm SUSY} \left(
\begin{array}{ccc}
  1    & (\delta_{U}^{12})_{LL} & (\delta_{U}^{13})_{LL}
\vspace{0.2cm} \\
(\delta_{U}^{21})_{LL} &    1   & (\delta_{U}^{23})_{LL}
\vspace{0.2cm} \\
(\delta_{U}^{31})_{LL} & (\delta_{U}^{32})_{LL} & 1
\end{array}\right),
\ee
and analogously for all the other blocks.

The related Feynman diagrams for $t\rightarrow cV \
(V=g,\gamma,Z)$ induced by the SUSY FCNC are shown in Fig.1.
Neglecting the charm quark mass, the amplitude of the decay
process is given by
\begin{eqnarray} M=\bar u(p_c)\, V^\mu  u(p_t)\, \epsilon_\mu(k,\lambda) \,,
\label{eq:deltaM} \end{eqnarray}where $p_t, p_c$, and $k$ are the
momenta of the top-quark, charm-quark, and gauge boson
respectively, and $\epsilon_\mu(k,\lambda)$ is the polarization
vector for the gauge boson. The vertex $V^\mu$ can be written as
\begin{eqnarray}
V^\mu &=&-i \gamma^\mu (P_L F_{V1}^L+ P_R F_{V1}^R) -i\frac{
p_t^\mu}{m_t}(P_L F_{V2}^L + P_R F_{V2}^R),\label{eq:Vtcv}
\end{eqnarray}
where $F_{V1(2)}^{L(R)}$ are the form factors, and their explicit
expressions through the SUSY-QCD FCNC
($\tilde{g}^a-\tilde{q}_i-q_I$) are:
\begin{eqnarray}
F_{g1}^L &=&\frac{-iT^a_{rs}}{96 \pi^2 m_t}\sum_{l=1}^6
\{m_{\tilde{g}} V_{7R}(9C_0^b m_t^2 V_{5L} V_6
+8V_4 V_{5R} B_0^f )+m_tV_{7L}[2V_{5L}C_{00}^dV_3+9V_6V_{5L} \nonumber\\
&& (2C_{00}^b+C_0^b(m_{\tilde{g}}^2-m_{\tilde{q}_l}^2)-C_2^bm_t^2-B_0^d)-8V_4V_{5R}B_1^f]-8m_{\tilde{g}}V_4V_{5L}V_{7R}B_0^e \},\\
F_{g1}^R&=&F_{g1}^L(V_{5L,R}\leftrightarrow
V_{5R,L},V_{7L,R}\leftrightarrow V_{7R,L}),\\
F_{g2}^L&=&\frac{-iT^a_{rs}}{48 \pi^2}\sum_{l=1}^6m_t
V_{5L}\{m_tV_{7L}[(C_{12}^d
+C_2^d+C_{22}^d)V_3+9(C_{12}^b+C_2^b+C_{22}^b)V_6]\nonumber\\
&&-m_{\tilde{g}} V_{7R}[(C_0^d+C_1^d+C_2^d)V_3-9(C_1^b+C_2^b)V_6]\},\\
F_{g2}^R&=&F_{g2}^L(V_{5L,R}\leftrightarrow
V_{5R,L},V_{7L,R}\leftrightarrow V_{7R,L}),\\
F_{\gamma1}^L&=&\frac{i\delta_{rs}}{12 \pi^2 m_t}\sum_{l=1}^6
[m_tV_{7L}(2C_{00}^dV_3V_{5L}
+V_4'V_{5R}B_1^f)+m_{\tilde{g}}V_4'V_{7R}(V_{5L}B_0^e
-V_{5R}B_0^f)],\\
F_{\gamma1}^R&=&F_{\gamma1}^L(V_{5L,R}\leftrightarrow
V_{5R,L},V_{7L,R}\leftrightarrow V_{7R,L}),\\
F_{\gamma2}^L&=&\frac{i\delta_{rs}}{6 \pi^2}\sum_{l=1}^6m_t
V_3'V_{5L}[m_t(C_{12}^d+C_2^d+C_{22}^d)V_{7L}-m_{\tilde{g}}(C_0^d+C_1^d+C_2^d)V_{7R}],\\
F_{\gamma2}^R&=&F_{\gamma2}^L(V_{5L,R}\leftrightarrow
V_{5R,L},V_{7L,R}\leftrightarrow V_{7R,L}),\\
F_{Z1}^L&=&\frac{i\delta_{rs}}{12 \pi^2
m_t}\sum_{l=1}^6[m_tV_{7L}(\sum_{l'=1}^62C_{00}^eV_3''V_{5L}+V_{4L}''V_{5R}B_1^f)\nonumber\\
&&+m_{\tilde{g}}V_{4L}''V_{7R}(V_{5L}B_0^e-V_{5R}B_0^f)],\\
F_{Z1}^R&=&F_{Z1}^L(V_{4L}''\rightarrow
V_{4R}'',V_{5L,R}\leftrightarrow
V_{5R,L},V_{7L,R}\leftrightarrow V_{7R,L}), \\
F_{Z2}^{L}&=&\frac{i\delta_{rs}}{6 \pi^2}\sum_{l,l'=1}^6m_t
V_3''V_{5L}[m_t(C_{12}^e+C_2^e+C_{22}^e)V_{7L}-m_{\tilde{g}}(C_0^e+C_1^e+C_2^e)V_{7R}],\\
F_{Z2}^{R}&=&F_{Z2}^{L}(V_{5L,R}\leftrightarrow
V_{5R,L},V_{7L,R}\leftrightarrow V_{7R,L}),
\end{eqnarray}
and the explicit expressions through the SUSY-EW FCNC
($\tilde{\chi}^0_k-\tilde{q}_i-q_I$) are:
\begin{eqnarray}
F_{g1}^L&=&\frac{iT^a}{16 \pi^2
m_t}\sum_{l=1}^6\sum_{k=1}^4 \{2C_{00}^am_tV_{1L}V_{2L}V_3+[(B^a_0m_{\tilde{\chi}^0_k}V_{1L} \nonumber\\
&&
+B_1^bm_tV_{1R})V_{2R}-B_0^bm_{\tilde{\chi}^0_k}V_{1R}V_{2L}]V_4
\},\\ F_{g1}^R&=&F_{g1}^L(V_{1L,R}\leftrightarrow
V_{1R,L},V_{2L,R}\leftrightarrow V_{2R,L}),\\
F_{g2}^L&=&\frac{iT^a}{8 \pi^2}\sum_{l=1}^6\sum_{k=1}^4m_t
V_{1L}V_3[(C_{12}^a+C_2^a+C_{22}^a)m_tV_{2L}-
(C_0^a+C_1^a+C_2^a)m_{\tilde{\chi}^0_k}V_{2R}],\\
F_{g2}^R&=&F_{g2}^L(V_{1L,R}\leftrightarrow
V_{1R,L},V_{2L,R}\leftrightarrow V_{2R,L}),\\
F_{\gamma1,2}^{L,R}&=&F_{g1,2}^{L,R}(V_3\rightarrow
V_3',V_4\rightarrow V_4', T^a\rightarrow1),\\
F_{Z1}^{L}&=&\frac{i}{16 \pi^2
m_t}\sum_{l=1}^6\sum_{k=1}^4\{\sum_{l'=1}^62C_{00}^fm_tV_{1L}V_{2L}V_3''
+V_{4L}''(-B_0^bm_{\tilde{\chi}^0_k}V_{1R}V_{2L}+B_1^bm_tV_{1R}V_{2R}\nonumber \\
&&+B^a_0m_{\tilde{\chi}^0_k}V_{1L}V_{2R})+\sum_{k'=1}^4m_tV_{1L}[(m_{\tilde{q}_l}^2V_{2L}V_{8R}
-m_{\tilde{\chi}^0_k}(m_tV_{2R}
+m_{\tilde{\chi}^0_{k'}}V_{2L})V_{8L})C_0^c
\nonumber\\
&& + V_{2L}V_{8R}B_0^c-2V_{2L}V_{8R}C_{00}^c
+(m_t^2V_{2L}V_{8R}+m_tm_{\tilde{\chi}^0_{k'}}V_{2R}V_{8R}\nonumber \\
&&-m_tm_{\tilde{\chi}^0_{k}}V_{2R}V_{8L})C_2^c]\},\\
F_{Z2}^L&=&\frac{i}{8
\pi^2}\sum_{l=1}^6\sum_{k=1}^4m_tV_{1L}\{\sum_{l'=1}^6V_3''[(C_{12}^a+C_2^a+C_{22}^a)m_tV_{2L}
-(C_0^a+C_1^a+C_2^a)m_{\tilde{\chi}^0_k}V_{2R}]\nonumber\\
&&+\sum_{k'=1}^4[-m_{\tilde{\chi}^0_k}V_{2R}V_{8L}C_1^c-V_{8R}(m_tV_{2L}C_{12}^c
+(m_tV_{2L}+m_{\tilde{\chi}^0_{k'}}V_{2R})C_{2}^c+\nonumber\\
&&m_tV_{2L}C_{22}^c)]\}, \\
F_{Z1,2}^{R}&=&F_{Z1,2}^{L}( V_{1L,R}\leftrightarrow
V_{1R,L},V_{2L,R}\leftrightarrow
V_{2R,L},V_{4L,R}''\leftrightarrow
V_{4R,L}'',V_{8L,R}\leftrightarrow V_{8R,L}).
\end{eqnarray}
Here $B^a_i = B_i(0, m_{\tilde{\chi}^0_k}^2, m_{\tilde{q}_l}^2)$,
$B^b_i = B_i(m_t^2, m_{\tilde{\chi}^0_k}^2,
m_{\tilde{q}_l}^2)$,$B^c_i = B_i(0, m_{\tilde{\chi}^0_k}^2,
m_{\tilde{\chi}^0_{k'}}^2)$, $B^d_i = B_i(0, m_{\tilde{g}}^2,
m_{\tilde{g}}^2)$,  $B^e_i = B_i(0, m_{\tilde{g}}^2,
m_{\tilde{q}_l}^2)$, $B^f_i = B_i(m_t^2, m_{\tilde{g}}^2,
m_{\tilde{q}_l}^2)$, $C^a_{i,ij} = C_{i,ij}(
0,0,m_t^2,m_{\tilde{\chi}^0_k}^2, m_{\tilde{q}_l}^2,
m_{\tilde{q}_l}^2)$,\\ $C^b_{i,ij} =
C_{i,ij}(0,0,m_t^2,m_{\tilde{q}_l}^2,
m_{\tilde{g}}^2,m_{\tilde{g}}^2)$, $C^c_{i,ij} = C_{i,ij}(
0,0,m_t^2, m_{\tilde{q}_l}^2,m_{\tilde{\chi}^0_k}^2,
m_{\tilde{\chi}^0_{k'}}^2)$, \\ $C^d_{i,ij} = C_{i,ij}(
0,0,m_t^2,m_{\tilde{g}}^2, m_{\tilde{q}_l}^2, m_{\tilde{q}_l}^2)$,
$C^e_{i,ij} = C_{i,ij}( 0,0,m_t^2,m_{\tilde{g}}^2,
m_{\tilde{q}_{l'}}^2, m_{\tilde{q}_{l'}}^2)$  \\and  $C^f_{i,ij} =
C_{i,ij}( 0,0,m_t^2,m_{\tilde{\chi}^0_k}^2, m_{\tilde{q}_{l}}^2,
m_{\tilde{q}_{l'}}^2)$ are 2 and 3-point one-loop integrals
\cite{loop}. And the relevant couplings are:
 \bea V_{1L}&=&
i\{\frac{-e}{\sqrt{2}s_Wc_W}(Z_U)_{2l}[\frac{1}{3}s_W(Z_N)_{1k}+c_W(Z_N)_{2k}]-Y_u^I(Z_U)_{5l}(Z_N)_{4k}\}, \nonumber\\
\ \
V_{1R}&=&i[\frac{2\sqrt{2}e}{3c_W}(Z_U)_{5l}(Z_N)_{1k}-Y_u^I(Z_U)_{2l}(Z_N)_{4k}], \nonumber\\
V_{2L}&=&
i\{\frac{-e}{\sqrt{2}s_Wc_W}(Z_U)_{3l}[\frac{1}{3}s_W(Z_N)_{1k}+c_W(Z_N)_{2k}]-Y_u^I(Z_U)_{6l}(Z_N)_{4k}\}, \nonumber\\
\ \
V_{2R}&=&i[\frac{2\sqrt{2}e}{3c_W}(Z_U)_{6l}(Z_N)_{1k}-Y_u^I(Z_U)_{3l}(Z_N)_{4k}], \nonumber\\
V_3&=& V_4=-ig_s, \ \ \ \ \ V_3'=V_4'=-i\frac{2}{3}e,
\nonumber\\V_3''&=&-i\frac{e}{2s_Wc_W}
[\sum_{I=1}^3(Z_U)_{Il}(Z_U)_{Il'}-\frac{4}{3}s^2_W\delta^{ll'}],\nonumber\\
V_{4L}''&=&-i\frac{e}{6s_Wc_W}(-3+4s^2_W),\ \ \
V_{4R}''=i\frac{2e s_W}{3c_W},\nonumber\\
V_{5L}&=&-i\sqrt{2}g_s(Z_U)_{2l},\ \ \
V_{5R}=i\sqrt{2}g_s(Z_U)_{5l},\ \ V_6=-ig_s,\nonumber\\
V_{7L}&=&-i\sqrt{2}g_s(Z_U)_{3l},\ \ \
V_{7R}=i\sqrt{2}g_s(Z_U)_{6l}, \nonumber\\
V_{8L}&=&i\frac{e}{2s_Wc_W}[(Z_N)_{4k}(Z_N)_{4k'}-(Z_N)_{3k}(Z_N)_{3k'}], \nonumber\\
\ \
V_{8R}&=&-i\frac{e}{2s_Wc_W}[(Z_N)_{4k}(Z_N)_{4k'}-(Z_N)_{3k}(Z_N)_{3k'}].
\nonumber\eea  After squaring the decay amplitude and multiplying
by the phase space factor, one obtains the decay width of
$t\rightarrow cV\ (V=g,\gamma, Z)$:
\begin{eqnarray}
&\Gamma (t \to cg,c\gamma) =  & \frac{1}{96 \pi}
m_{t}[(2F_{V1}^R-F_{V1}^L)F_{V1}^{L\ast}+(2F_{V1}^L-F_{V2}^R)F_{V1}^{R\ast}-(F_{V1}^L+F_{V2}^R)F_{V2}^{L\ast}\nonumber
\\&&-(F_{V1}^R+F_{V2}^L)F_{V2}^{R\ast}],
 \\
& \Gamma (t \to cZ) = & \frac{1}{384 \pi m_Z^2m_t^5} (m_t^2-m_Z^2)^2\{2m_t^2F_{Z1}^{L\ast}[m_Z^2(4F_{Z1}^R-F_{Z2}^L)+m_t^2(2F_{Z1}^R+F_{Z2}^L)]\nonumber \\
&&+2m_t^2F_{Z1}^{R\ast}[m_Z^2(4F_{Z1}^L-F_{Z2}^R)+m_t^2(2F_{Z1}^L+F_{Z2}^R)]-(m_t^2-m_Z^2)[F_{Z2}^{L\ast}\nonumber\\
&&(m_Z^2F_{Z2}^R-m_t^2(2F_{Z1}^L+F_{Z2}^R))+F_{Z2}^{R\ast}(m_Z^2F_{Z2}^L-m_t^2(2F_{Z1}^R+F_{Z2}^L))]\},
 \end{eqnarray}and we define the branching ratio as
 Ref.~\cite{eilam}:
\begin{eqnarray}
Br(t\rightarrow cV)\equiv\frac{ \Gamma(t\rightarrow cV)}{
\Gamma(t\rightarrow bW^+)},
\end{eqnarray}
which will be the main object of our numerical study.
\section{\label{sec:dis}Numerical calculation and discussion}
In our numerical calculations the SM parameters were taken to be
$m_t=174.3$ GeV, $M_W=80.423$ GeV, $M_Z=91.1876$ GeV,
$\sin^2\theta_W=0.23113$ and $\alpha_s(M_Z)=0.1172$ \cite{pdg}.
The relevant SUSY parameters are $\mu$, $\tan\beta$, $M_{\rm
SUSY}$ and $m_{\tilde{g}}$, which are unrelated to flavor changing
mechanism, and may be fixed from flavor conserving observables at
the future colliders. And they are chosen as follows:
$M_{\rm{SUSY}}=400, 1000$ GeV, $\tan \beta =4, 40$,
$m_{\tilde{g}}=200, 300$ GeV and $\mu = 200$ GeV. As for the range
of the flavor mixing parameters, $(\delta^{ij}_{U})_{LL}$ are
constrained by corresponding
$(\delta^{ij}_{D})_{LL}$\cite{gam,besmer,gabre,saha}, in which
$(\delta^{12}_{U})_{LL}$ also is constrained by the chargino
contributions to $K$-$\bar{K}$ mixing\cite{leb}, and
$D_0$-$\bar{D_0}$ mixing makes constraints on
$(\delta^{12}_{U})_{LL}$, $(\delta^{12}_{U})_{LR}$ and
$(\delta^{12}_{U})_{RL}$\cite{wyler}. And
$(\delta^{31}_{U})_{LL}$, $(\delta^{32}_{U})_{LL}$,
$(\delta^{31}_{U})_{RL}$ and $(\delta^{32}_{U})_{RL}$ are
constrained by the chargino contributions to $B_d$-$\bar{B_d}$
mixing\cite{gabre}. Finally, there also are constraints on the up
squark mass matrix from the chargino contributions to
$b\rightarrow s\gamma$\cite{causse, besmer}. Taking into account
above constraints, in our numerical calculations, we use the
following limits:

(i) $(\delta^{12}_{U})_{LL}$, $(\delta^{12}_{U})_{LR}$ and
$(\delta^{12}_{U})_{RL}$ is less than $0.08M_{\rm SUSY}$/(1 TeV);

(ii) $(\delta^{12}_{U})_{RR}$ and $(\delta^{13}_{U})_{LL}$ are
limited below $0.2M_{\rm SUSY}$/(1 TeV);

(iii) $(\delta^{23}_{U})_{LL}$, $(\delta^{23}_{U})_{LR}$,
$(\delta^{23}_{U})_{RL}$, $(\delta^{23}_{U})_{RR}$,
$(\delta^{13}_{U})_{LR}$, $(\delta^{13}_{U})_{RL}$ and
$(\delta^{13}_{U})_{RR}$ vary from 0 to 1.

First of all we should point out that the contributions arising
from SUSY-EW FCNC are in general at least one magnitude of order
smaller than ones arising from SUSY-QCD FCNC, and the dominant
contributions to the decay branching ratios come from the latter.
Furthermore, our calculations show that the decay branching ratios
only weakly depends on $\tan \beta$, so we only discuss the
results in the case of $\tan \beta=40$ below. Our results are
shown in Figs.~\ref{fig:tcg}--\ref{fig:last}, where there are
three common features of these curves: the first is that the
branching ratio increases rapidly with the mixing parameters
increasing, and the second is that the branching ratio depends
strongly on the gluino mass $m_{\tilde{g}}$, and the last is that
the dependence of branching ratio on the $M_{\rm SUSY}$ is medium
comparing with above two parameters.

For each decay modes $t\rightarrow cV\ (V=g,\gamma, Z)$, in
Figs.~\ref{fig:tcg} and \ref{fig:last} we show the dependence of
the decay branching ratios on RR and LR off-diagonal elements,
respectively (We do not show the results for LL off-diagonal
elements as their contributions are similar to the ones for RR
off-diagonal elements). We find that for $t \rightarrow cV$ the
largest results come from the LR block, which arises from the soft
trilinear couplings $A_{U}$. We also give the results of the
interference effects on the branching ratios for $t\rightarrow cg$
in Figs.~\ref{fig:tcg}(c)and \ref{fig:tcg}(d). In general, these
interference effects increase the decay branching ratios, and
since the interference effects between different blocks are
similar, we do not show them for the space of this paper. The
results of decay $t \rightarrow c\gamma, cZ$ are about two orders
of magnitude smaller than ones of decay $t \rightarrow cg$, as
shown in Figs.~\ref{fig:tcg} and \ref{fig:last}. From these
figures, we can find that the decay branching ratios for $t
\rightarrow cV \ (V=g,\gamma, Z)$ induced by the SUSY FCNC
couplings can reach $\sim10^{-4}$, $\sim10^{-6}$ and
$\sim10^{-6}$, respectively, for the favorable parameter values
allowed by current precise experiments, and they are larger than
all the previous ones in the MSSM with R-parity conservation (it
should be pointed out that the results of Ref.~\cite{csldecay},
\cite{cou1} and \cite{cou2} in Table 1 are obtained at
$m_{\tilde{g}}=100$ GeV, which is disfavored by current data).

According to the analysis of T.~Han {\it et al.} \cite{han}, the
sensitivities for $t\rightarrow c\gamma$ and $t\rightarrow cZ$ at
the LHC with $100 fb^{-1}$ integrated luminosity are $5\times
10^{-6}$ and $2\times 10^{-4}$, respectively, and our results show
that the rare decay $t\rightarrow c\gamma$ may be detectable.
Later T.~Han {\it et al.} \cite{han2} and M.~Hosch {\it et
al.}~\cite{hosch} studied the sensitivities to the top quark
anomalous FCNC couplings at the LHC for single top quark and
direct top quark productions, respectively, and the corresponding
decay $t\rightarrow cg$ branching ratios transferred from their
results are $4.9\times10^{-5}$ and $2.7\times10^{-5}$,
respectively. Thus, our results of the branching ratios for
$t\rightarrow cg$ indicate that the top quark FCNC production
processes (both for single top and direct top) may be measurable
at the LHC. But if we use the $5\times10^{-5}$ as the sensitivity
for the FCNC decay $t\rightarrow cV$ at the LHC with $100 fb^{-1}$
integrated luminosity as shown in Ref.~\cite{sola}, our results
show that the rare decay $t\rightarrow cg$ are also potentially
measurable at the LHC.

In conclusion, we have calculated the top quark rare decay
$t\rightarrow cV$ ($V=g,\gamma,Z)$ induced by SUSY-FCNC couplings
in the general unconstrained MSSM using mass eigenstate approach.
Our results show that the branching ratios for these decays are
larger than ones reported in previous literatures in the MSSM with
R-parity conservation, and especially, the branching ratios for
the rare decay modes $t\rightarrow cg,c\gamma$ we calculated are
very hopefully to be measurable at the LHC for the favorable
parameter values allowed by current precise experiments. Moreover,
we find that the decay branching ratios for $t\rightarrow
cV(V=g,\gamma,Z)$ strongly depend on the soft trilinear couplings
$A_{U}$, and it is possible to get some valuable information of
soft SUSY breaking parameters by measuring the branching ratio for
the top quark rare decay at the LHC.

\vspace{3.5cm}

\hspace*{4.5cm}ACKNOWLEDGMENTS\\

We would like to thank T. Han for useful discussions. This work
was supported in part by the National Natural Science Foundation
of China and Specialized Research Fund for the Doctoral Program of
Higher Education.

\vspace{1cm}


\newpage
\begin{figure}
\centerline{\epsfig{file=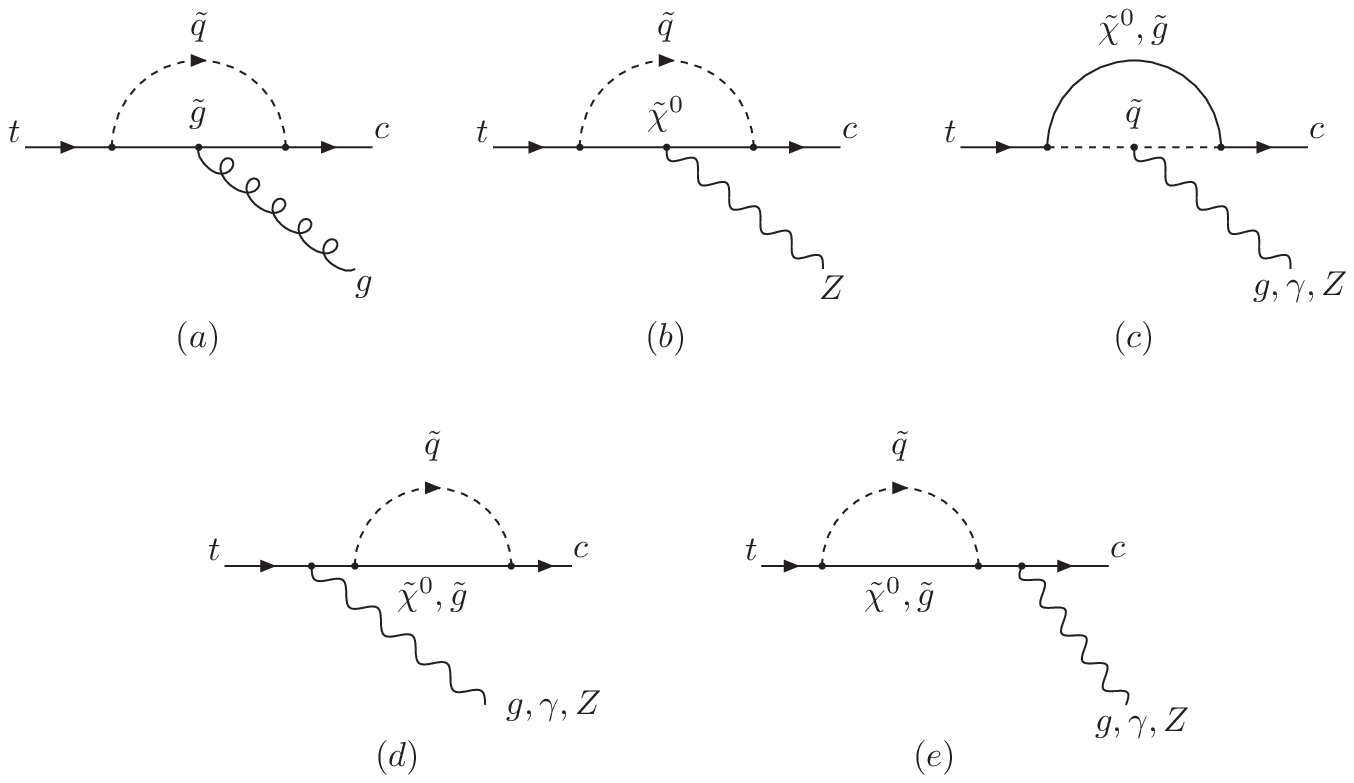,width=400pt}}
\caption[]{Feynmann diagrams for $t \rightarrow cV(V=g,\gamma,Z)$.
\label{fig:feyn}}
\end{figure}

\newpage

\begin{figure}
\centerline{\epsfig{file=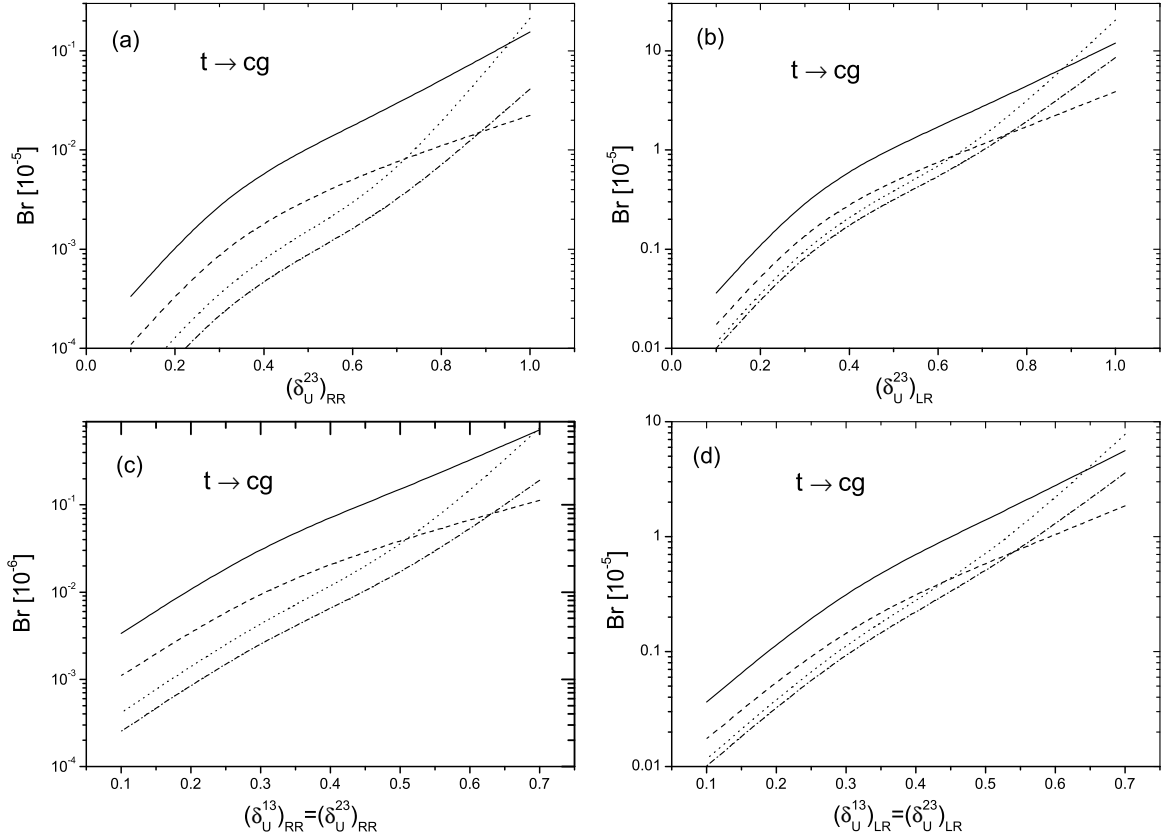, width=500pt}} \caption[]{The
decay branching ratios for the $t\rightarrow c g$ with mixed RR
off-diagonal elements(a) and LR off-diagonal elements(b), and the
typical interference effects of RR block(c) and LR block(d). Here,
solid line: $m_{\tilde{g}}=200$ GeV, $M_{\rm SUSY}=400$ GeV;
dashed line: $m_{\tilde{g}}=300$ GeV, $M_{\rm SUSY}=400$ GeV;
dotted line: $m_{\tilde{g}}=200$ GeV, $M_{\rm SUSY}=1000$ GeV;
dash-dotted line: $m_{\tilde{g}}=300$ GeV, $M_{\rm SUSY}=1000$
GeV. \label{fig:tcg}}
\end{figure}

\newpage

\begin{figure}
\centerline{\epsfig{file=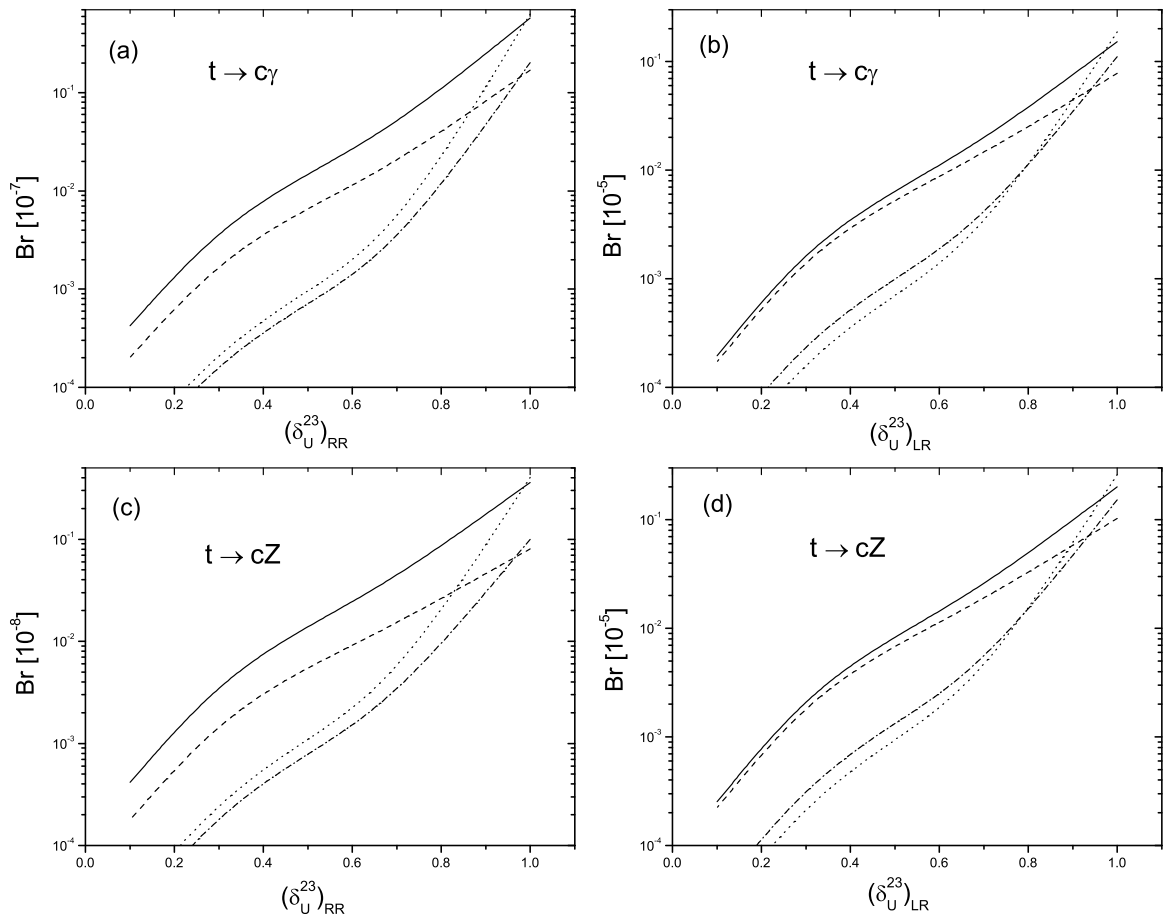, width=500pt}} \caption[]{The
decay branching ratios for the $t\rightarrow c \gamma$ with mixed
RR off-diagonal elements(a) and LR off-diagonal elements(b), and
the decay branching ratios for the $t\rightarrow c Z$ with mixed
RR off-diagonal elements(c) and LR off-diagonal elements(d). Here,
solid line: $m_{\tilde{g}}=200$ GeV, $M_{\rm SUSY}=400$ GeV;
dashed line: $m_{\tilde{g}}=300$ GeV, $M_{\rm SUSY}=400$ GeV;
dotted line: $m_{\tilde{g}}=200$ GeV, $M_{\rm SUSY}=1000$ GeV;
dash-dotted line: $m_{\tilde{g}}=300$ GeV, $M_{\rm SUSY}=1000$
GeV. \label{fig:last}}
\end{figure}

\newpage

\end{document}